# Seamless design of information system architecture based on adaptive clustering method

*Grigory Tsiperman*
*Assistant professor*
*National University of Science and Technology "MISIS"*
*gntsip@gmail.com*

**Abstract**
  The paper considers the concept of building the architecture of an information system that provides a seamless connection between architectural representations of various levels of abstraction. The concept is based on the application of the adaptive clustering method of information systems developed by the author. Seamless connection is understood as the presence of connections between elements of architectural models related to architectural representations of various levels of abstraction.

## 1 Introduction

  This paper considers consider the issues of building an information system (IS) architecture in the context of enterprise architecture. The concept of IS architecture historically preceded the concept of enterprise architecture, which arose from the realization that the IS model must meet the requirements of the business and be able to flexibly adapt to its needs. Moreover, such compliance should not be fragmented, it should be based on a deep understanding of the business and its development prospects. IS should ensure the implementation of business requirements, have the ability to adapt to its changes.

  The idea of enterprise architecture is to create interconnected architectural models that combine the concepts of mission, goals, enterprise business strategy, business processes, information systems, etc. To implement the idea of enterprise architecture, several methodologies were proposed (see, for example, the reviews in [1] and [2]), none of which are without drawbacks and to this day is not a paradigm.

  What all the methodologies agree on is the need for the concept of the life cycle of an enterprise and, accordingly, IS. For the TOGAF [3], GERAM [4] and FEA [5] ICs, the identical stages of the life cycle and the types of IS architecture corresponding to these stages are determined:

- Business architecture
- IS architectures (data architecture and application architecture)
- Architecture of technology.

  This paper discusses the construction of IS architecture, so the IS model under consideration includes only the necessary aspects of the enterprise business process model. Moreover, the work does not concern the construction of a basic (as-is) model, confining itself to the issues of constructing a target (to-be) architecture of IS based on a target business model.

One way or another, starting with Zakhman [6], the methodologies for creating IS architecture suggest a downward development, with the consistent construction of architectural models of an ever-increasing level of detail. The levels of detail correspond to certain stages of the IS life cycle: the concept is created by presenting the target business architecture, technical design involves the description of functional architecture, component architecture, data architecture, and determines the technology architecture of IS.

The practice of designing IS architecture involves a heuristic transition between architectural models of various levels of abstraction: a more detailed architecture is constructed in such a way that the technical solutions match the models of the previous level as much as possible. In this case, the architect uses his creativity, experience and knowledge to build a detailed model, and then proves (or considers it obvious) that the resulting model satisfies the requirements arising from a more abstract architecture. In the heuristic approach, the relationship between architectural models of various levels of abstraction is inverse evidence-based. Many developers paid attention to this aspect of IS design (see [7], [8], [9]). Let us designate it as a technological gap between architectural models of various levels of abstraction.

In fact, we are talking about verification and validation[1] the architectural models that underlie the program code. Technological gaps in the absence of validation and validation processes can lead to costly errors.

Unlike the heuristic transition between architectural models of various levels of abstraction, which creates technological gaps, this article considers a smooth transition in which such gaps do not arise. The technology under consideration relates to the development of functional IS architectures. It is based on the application of the adaptive clustering method (ACM, see [10], [11]), in which detailed architectural models are justified by high-level abstraction models, ensuring their seamless connection and the possibility of tracing between components of architectural models.

## 2 Formulation of the problem

The architectural description of the system is a set of architectural representations corresponding to various points of view on the system. Consider Figure 1, representing a fragment of a conceptual model of architectural representation [12], supplemented by an architectural element that defines a seamless connection between architectural representations.

The architectural representation corresponds to a certain point of view on architecture, and defines the architecture of IS with a degree of detail corresponding to the stakeholder. In Zakhman's scheme, for example, the points of view correspond to the participants in the process of creating the system: a planner, analyst, architect, designer, programmer and operator. Considered in the indicated sequence, these points of view lead to associated architectural representations of an ever-increasing level of detail. An architectural representation includes one or more architectural models that reflect the relationship of the elements of the architecture description (hereinafter referred to as the elements). Elements correspond to abstracts determined by the type of model and the point of view on the architecture of IS.

The main objective of this work is to identify among the elements of architectural models included in the architectural representation of the corresponding level of abstraction, an element (connecting element), the decomposition of which defines the elements of architectural models of the next level of abstraction.

---

[1] Validation is a confirmation (based on the presentation of objective evidence) that the requirements intended for a particular use or application are fulfilled, and verification is a confirmation that the specified requirements are fulfilled [15].

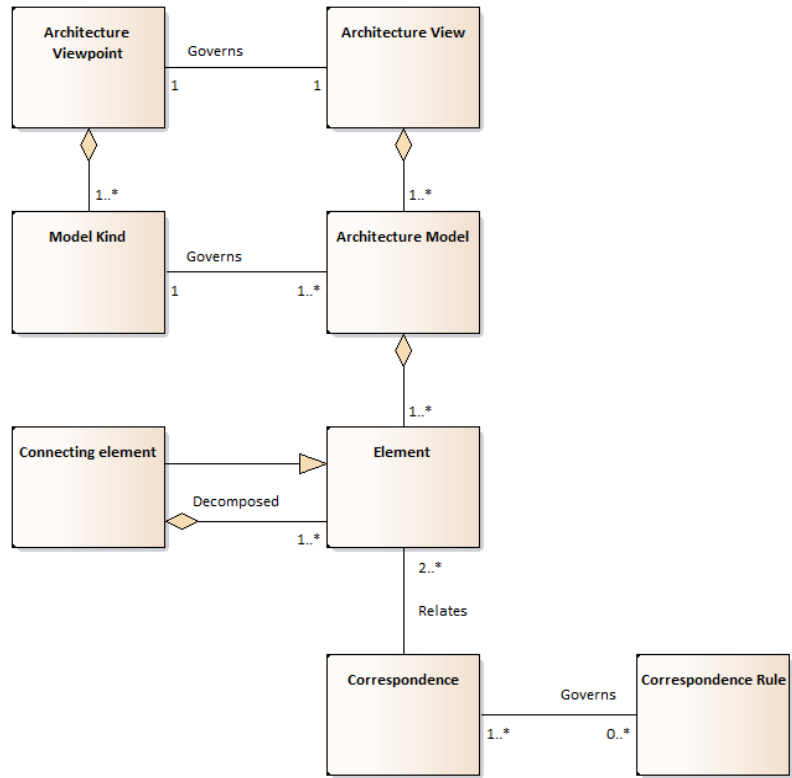

Figure 1. Conceptual model of architectural representation

The connecting element clearly defines the relationship between the elements of architectural models of various architectural representations. The essence of the technological gap between architectural concepts is that with a heuristic approach to design, the relationships between the elements of architectural models are not defined explicitly, so the validation and verification of architectural models is a rather complicated, painstaking task.

By seamless connection, we mean the presence of explicit connections between the elements of architectural models of various architectural representations. In this case, the description of the IS architecture is an interconnected set of architectural representations, and for elements of architectural models, tracing[2] becomes possible, which makes the process of model validation and verification trivial - you just need to make sure that the decomposition of the connecting elements is performed correctly.

In this paper, based on the definition of connection elements, describe a seamless connection scheme of architectural representations of the target business architecture, functional architecture, component architectures, data architecture and deploying architecture IS.

## 3 Business Architecture Design

At ACM, it is business architecture that is the starting point for IS design. In the context of IS design, business architecture is understood to mean many models of automated business processes of an enterprise. In the context of this work, we consider the target business architecture, i.e. business process models taking into account the use of IS. The task is to determine the complete set of automated functions of the business process.

---

[2] The relationship between the elements of architectural models different levels of abstraction.

## 3.1 Business Architecture Concepts

The architectural models of business architecture are built on three main elements (Figure 2), which are defined in the decomposition procedure:

- a business process, understood in the usual sense, as an action in which, based on one or more types of input objects [resources], a result valuable to the client is created; a business process can be represented by a set of actions determined on the basis of its decomposition;
- a business function is an action from a set of actions determined by the decomposition of a business process;
- a business operation is a business function that cannot be decomposed, the executor of which is a specific employee.

Decomposition determines the explicit relationships between these architectural elements: a business process is decomposed into business functions, business functions, in turn, are decomposed into business functions and business operations. The process ends when all business functions are decomposed into business operations.

A business operation is a place where a user interacts with an IS, and accordingly, explicitly or implicitly, includes automated functions that support it.

## 3.2 Operational Service

The business architecture defines the functional requirements of the user for IS. Description of business operations allows you to define functions that should be automated. The set of these functions, defined for a business operation, constitutes the content of the business operation service (hereinafter referred to as the operational service). Operational services are formed for each business operation and include automated functions descriptions. They are architectural elements of an abstract nature that do not imply any implementation.

An operational service is a connecting element that defines a seamless relationship between business architecture and functional architecture. Figure 2 demonstrates this relationship: the operational service is decomposed into system dialogs, which are the main element of the functional architecture that provides access to IS functions. Automated functions of a business operation are implemented in dialogs.

In addition, the operational service is the boundary between the business architecture and the system architecture. It formalizes the requirements for automating business operations and serves as the basis for the formation of software requirements specifications.

## 4 Functional Architecture

Functional architecture is understood as an architectural representation, including architectural models of the structure and composition of the functional components of IS, providing access to the "internal" functions of IS that implement automated functions. In other words, the functional architecture models the interaction of IS with users, as well as with other external agents. The functional architecture forms the appearance of the IS based on the presentation of the compositions of the dialogs of the system, defines the requirements for dialogs and specifies interfaces with external systems.

A dialogue is any act of agent interaction that causes a change in the state of the IS by launching the corresponding software components [9]. Thus, dialogue is understood in a broad sense - this is not only the interaction of the user with the computer, but also the exchange of messages between any IS objects, as well as external agents.

The structure of the dialogue description (Figure 2) includes the following architectural elements:

- The function of the presentation level of the IS (view function) is the IS function implemented in the dialogue, providing access to the "internal" IS functions. In other words, through the view functions, the user interacts with the IS.
- The source resource and the Target product are information objects, respectively, received at the input of the dialogue upon its initiation and formed as a result of the dialogue at the exit from it.
- Dialog form - representing a dialog for the user, if the user is supposed to be an agent.

The view functions defined in the dialog have the following classification:

- implementation of dialogue preconditions upon initiation;
- input / output of data field values;
- processing of dialogue control elements that change the state of the IS;
- reaction to errors and contingencies during the dialogue;
- implementation of postconditions at the end of the dialogue.

Functional architecture defines the next level of architectural representation - component architecture. A connecting element that defines seamless connections between functional architecture and component architecture are view functions that are decomposed into software modules. A detailed discussion of component architecture is provided in the next section.

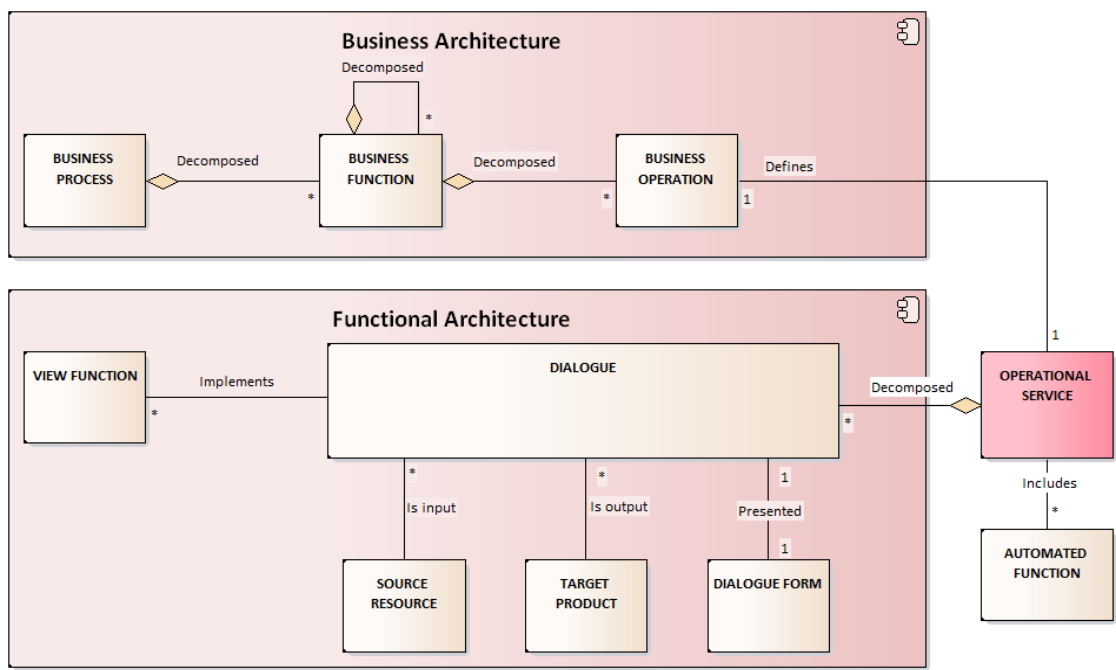

Figure 2. Connection between business architecture and functional architecture

## 5 Component Architecture

The view functions resulting from the design of dialogs do not take into account the component architecture of the system. These functions concern only the presentation level of the IS. Representation of the IS architecture in the form of interacting functional components (subsystems and external systems) makes it possible to find out how the view functions implement, to determine the internal functions of the application logic and data management.

The component architecture establishes the composition and interaction of the functional components of the IS, determines the software modules and their distribution according to the functional components, details the functional requirements for the IS.

The main elements of the component architecture model are (Figure 3):

- functional components (subsystems and external systems), determined by the choice of an architectural template (or a composition of architectural templates) and the external environment of the system;
- software modules, which are the structural parts of the functional components of the IS and are determined by the decomposition of the view function.

The design of component architecture begins with the definition of the component structure of the IS and of which represent the system as the set functional components (subsystems and external systems). The modular composition of the functional components of the system is determined by the decomposition of the connecting elements of the functional architecture - the view functions of IS, performed on the selected component structure. The decomposition of the view function may include previously defined modules for reuse, i.e. there is a many-to-many relationship between view functions and system modules.

In ACM, to determine the modular composition of IS, Sequence Diagrams are used, which are generated for the view functions of each dialogue described at the level of functional architecture. Functional components are used as lifelines in diagrams. Thus, design allows you to define a complete set of software modules for all functional components.

## 6 Definition of class methods

ACM offers a technology for developing a data architecture (see [13] and [14]), but an exposition of this technology is beyond the scope of this work. We proceed from the fact that the data architecture in one way or another is developed at the level of the ER model and includes all the necessary attributes, and the entities are distributed among the elements of the component model. The task is to determine the data architecture classes methods based on the generated modular structure: for each module of the component model, the order of its implementation within the framework of the object-oriented IS model is determined.

Thus, the connecting element that determines the seamless transition from component architecture to data architecture are the software modules of the component model. Class methods are determined by the decomposition of each module on the presented data architecture (Figure 3). As with the component model, there is a many-to-many relationship between modules and class methods, corresponding to the reuse of methods to implement modules. Class methods are also defined using Sequence Diagrams in which the lifelines correspond to the classes of the data model.

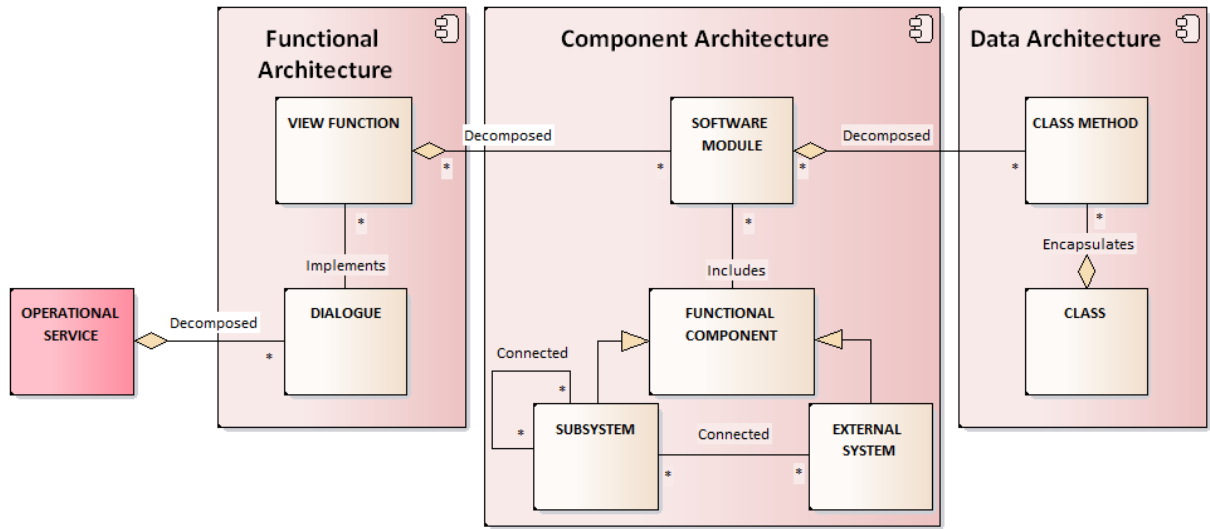

Figure 3. Functional architecture, component architecture, data architecture and their relationship

## 7 Technological Architecture

Technological architecture is a model representing the technical infrastructure of IS, including solutions in the field of computing and telecommunications infrastructure.

Technological architecture, in the context of this work, represents the distribution of elements of component architecture (deployment) over various hardware and determines the necessary interfaces between objects of technical architecture. The choice of architectural solutions for technological architecture is limited by system requirements, quality attributes, and requirements for external interfaces.

In terms of level of abstraction, technological architecture is superior to component architecture. Figure 4 shows that the connecting element between these architectures is the element of the technological architecture corresponding to the hardware. Connection reflect the location of functional components on the IS hardware.

Architectural representations of technological architecture are typically performed using Deployment Diagrams. The description of the deployment diagram objects may include either requirements for the characteristics of the hardware-software tool, or an indication of a specific device model.

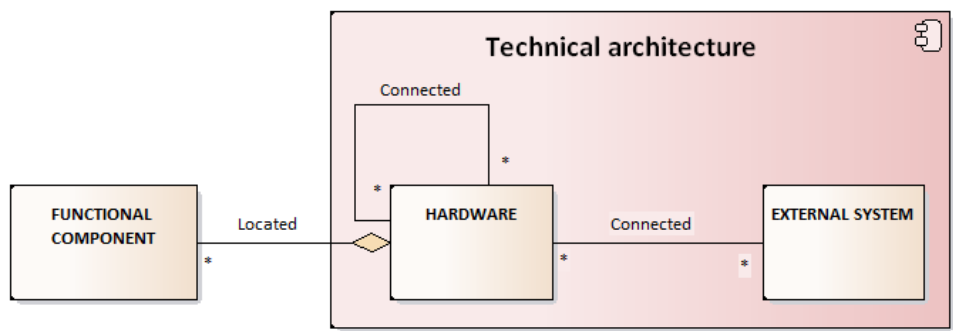

Figure 4. The relationship of component architecture and technical architecture

# 8 Conclusion

The method considered in this work provides a seamless transition from business architecture through the decomposition of the operational services to the functional architecture of the IS, which defines the dialogs of the system and the view functions. The decomposition of the view functions defines the software modules of the component architecture and, finally, the modules are decomposed into classes methods of the data architecture. In other words, each architectural representation is derived from the architecture of the previous level of abstraction.

With this approach, the completeness of the functional implementation of the IS is ensured, since the decomposition of the business process allows you to accurately formulate the user requirements for the automated functions. Further design of functional components at various levels of the abstract description of IS is essentially a reasonable conclusion of the necessary and sufficient functionality of the IS, which avoids errors associated with technological gaps between architectural models, insufficient or excessive functionality.

The advantages shown by the ACM in real-world design and maintenance of IS include ensuring transparency of the compliance of the design result with the requirements set by the customer by simplifying the validation and verification processes. The presence of tracing between the elements of architectural models allows for the rapid localization of necessary changes and improvements for the release of new versions of IS. The regulation of the development process, the relationship of the architectural description with artifacts, taking into account the possibility of generating design and operational documents based on architectural models [11], provides a significant reduction in the design time and making necessary changes to IS.

In conclusion, I would like to outline topics that were not reflected in this work. Further research is required by the relationship between ACM and service-oriented architecture (SOA), namely the identification of services and the implementation of the component architecture of IC on the SOA template. In general, working with external information resources, whether it's web services or library classes, requires study in the context of the ACM.

**Acknowledgments**

The author expresses sincere gratitude to the professor, Doctor Boris Pozin, who provided invaluable assistance in the preparation of this article.

# Бесшовное проектирование архитектуры информационной системы на основе метода адаптивной кластеризации

*Циперман Григорий Наумович*
*Доцент кафедры системной и программной инженерии МИСиС*
*117485, Москва, ул. Профсоюзная, д. 102/47, кв. 87*
*gntsip@gmail.com*

**Аннотация**
В работе рассматривается концепция построения архитектуры информационной системы, обеспечивающая бесшовную связь между архитектурными представлениями различного уровня абстракции. Концепция основана на применении разработанного автором метода адаптивной кластеризации информационных систем. Бесшовная связь, понимается как наличие связей между элементами архитектурных моделей, относящихся к архитектурным представлениям различных уровней абстракции.

**Ключевые слова**: информационная система, архитектурная абстракция, проектирование, архитектурная модель, бесшовная связь, технологический разрыв, метод адаптивной кластеризации

## 1 Введение

В данной работе мы рассмотрим вопросы построения архитектуры информационной системы (ИС) в контексте архитектуры предприятия. Понятие архитектуры ИС исторически предшествовало понятию архитектуры предприятия, которое возникло из осознания того, что модель ИС должна соответствовать требованиям бизнеса и быть способна гибко подстраиваться по его нужды. Причем, такое соответствие не должно носить фрагментарный характер, оно должно быть основано на глубоком понимании бизнеса и перспектив его развития. ИС должна обеспечивать реализацию требований бизнеса, обладать возможностями адаптации к его изменениям.

Идея архитектуры предприятия заключается в создании взаимосвязанных архитектурных моделей, объединяющих понятия миссии, целей, бизнес-стратегии предприятия, бизнес-процессов, информационных систем и т.д. Для реализации идеи архитектуры предприятия было предложено несколько методологий (см., например, обзоры в [15] и [16]), ни одна из которых не лишена недостатков и к сегодняшнему дню не является парадигмой. Однако, следует отметить, что методология GERAM [3] легла в основу отечественного ГОСТ Р ИСО 15704-2008 [4].

В чем согласны все методологии, так это в необходимости понятия жизненного цикла предприятия и, соответственно ИС. Для ИС TOGAF [5] и FEA [6] определяют одинаковые стадии жизненного цикла и соответствующие этим стадиям типы архитектуры ИС:

– Архитектура бизнеса
– Архитектуры ИС (архитектура данных и архитектура приложений)
– Архитектура технологии.

В настоящей работе рассматриваются вопросы построения архитектуры ИС, поэтому рассматриваемая модель ИС включает только необходимые аспекты модели бизнес-процессов предприятия. Более того, работа не касается построения базовой (as-is) модели, ограничиваясь вопросами построения целевой (to-be) архитектуры ИС, основанной на целевой бизнес-модели.

Так или иначе, начиная с Захмана [7], методологии создания архитектуры ИС, предполагают нисходящую разработку, с последовательным построением архитектурных моделей все большего уровня детализации. Уровни детализации соответствуют определенным стадиям и этапам жизненного цикла ИС: созданию концепции соответствует представление целевой бизнес-архитектуры, технорабочее проектирование предполагает описания функциональной, компонентной архитектур, архитектуры данных, и определяет технологическую архитектуру ИС.

Практика проектирования архитектуры ИС предполагает эвристический переход между архитектурными моделями различных уровней абстракции: более детальная архитектура строится таким образом, чтобы технические решения максимально соответствовали модели предыдущего уровня. При этом архитектор использует свой креатив, опыт и знания для построения детальной модели, а затем доказывает (или считает это очевидным), что полученная модель удовлетворяет требованиям, вытекающим из более абстрактной архитектуры. При эвристическом подходе связь между архитектурными моделями различных уровней абстракции имеет обратный доказательный характер. На этот аспект проектирования ИС обращали внимание многие разработчики (см. [8], [18], [19]). Обозначим его как технологический разрыв между архитектурными моделями различных уровней абстракции.

По сути, речь идет о верификации и валидации[3] архитектурных моделей, лежащих в основе программного кода. Технологические разрывы при отсутствии процессов верификации и валидации могут приводить к дорогостоящим ошибкам.

В противоположность эвристическому переходу между архитектурными моделями различных уровней абстракции, порождающему технологические разрывы, в настоящей работе рассматривается бесшовный переход, в котором такие разрывы не возникают. Рассматриваемая технология касается проектирования функциональных архитектур ИС. Она основана на применении метода адаптивной кластеризации (МАК, см. [20], [21]), в котором детальные архитектурные модели строятся на основе вывода из моделей высокого уровня абстракции, обеспечивая их бесшовную связь и возможность трассировки между компонентами архитектурных моделей.

**2 Постановка задачи**

Архитектурное описание системы представляет собой совокупность архитектурных представлений, соответствующих различным точкам зрения на систему. Рассмотрим Рисунок 1, представляющий фрагмент концептуальной модели архитектурного представления [22], дополненный архитектурным элементом, определяющим бесшовную связь между архитектурными представлениями.

Архитектурное представление соответствует определенной точке зрения на архитектуру, и определяет архитектуру ИС со степенью подробности, соответствующей интересам заинтересованного лица. В схеме Захмана, например, точки зрения соответствуют участникам процесса создания системы: планировщик, аналитик, архитектор, проектировщик, программист и оператор. Рассматриваемые в указанной последовательности эти точки зрения приводят к связанным архитектурным представлениям все большего уровня детализации. Архитектурное представление включает одну или несколько архитектурных моделей, отражающих взаимосвязи элементов описания архитектуры (далее – элементы). Элементы соответствуют абстрактам, определяемым видом модели и точкой зрения на архитектуру ИС.

---

[3] Валидация – это подтверждение (на основе представления объективных свидетельств) того, что требования, предназначенные для конкретного использования или применения, выполнены, а верификация – подтверждение того, что заданные требования выполнены [16].

Основной задачей настоящей работы является выявление среди элементов архитектурных моделей, входящих в архитектурное представление соответствующего уровня абстракции, элемента (связующего элемента), декомпозиция которого определяет элементы архитектурных моделей следующего уровня абстракции, детализирующего предыдущий.

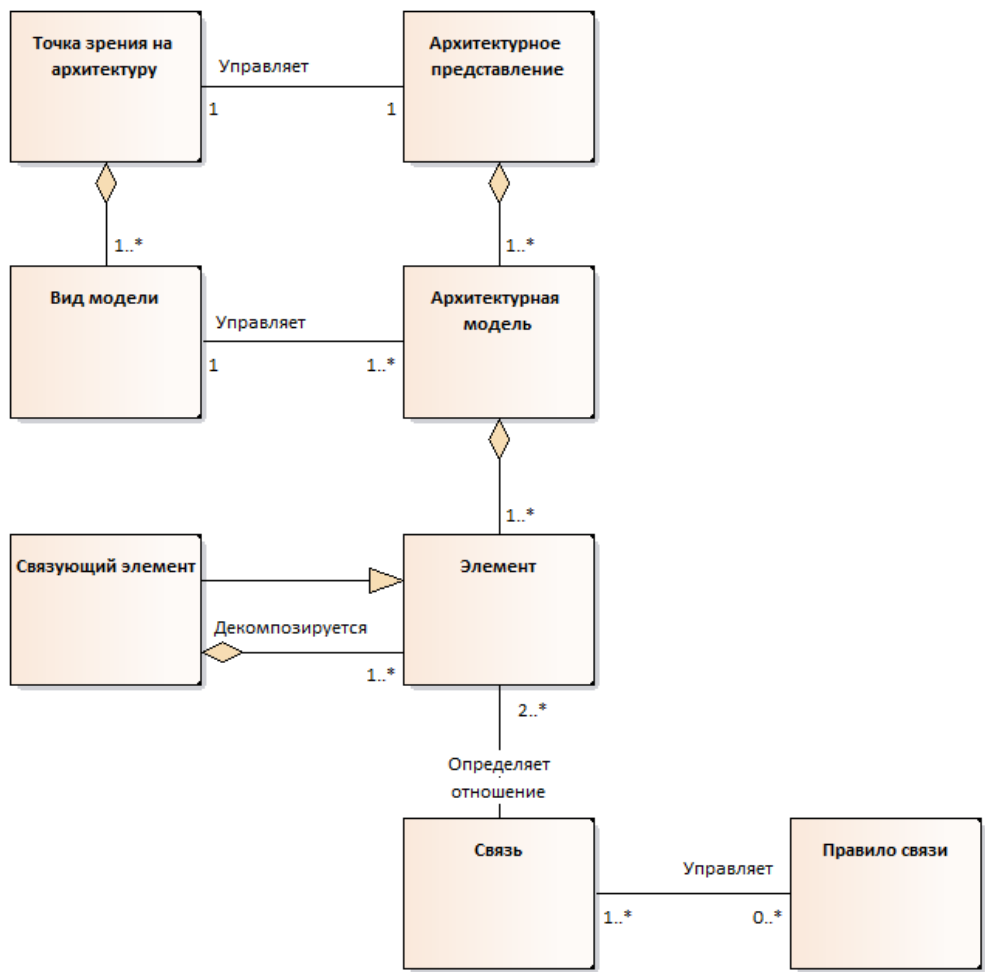

Рисунок 1 - Концептуальная модель архитектурного представления

Связующий элемент явно определяет связи между элементами архитектурных моделей различных архитектурных представлений. Суть технологического разрыва между архитектурными представлениями заключается в том, что при эвристическом подходе к проектированию, связи между элементами архитектурных моделей не определяются явно, поэтому валидация и верификация архитектурных моделей представляют собой достаточно сложную, кропотливую задачу.

Под бесшовной связью мы будем понимать наличие явных связей, между элементами архитектурных моделей различных архитектурных представлений. В этом случае описание архитектуры ИС представляет собой взаимоувязанное множество архитектурных представлений, а для элементов архитектурных моделей становится возможна трассировка[4], делающая тривиальным процесс валидации и верификации моделей – просто надо убедиться, что декомпозиция связующего элемента выполнена правильно.

В настоящей работе на основе определения связующих элементов описана схема бесшовной связи архитектурных представлений целевой бизнес-архитектуры, функциональной и компонентной архитектур, архитектуры данных и архитектуры развертывания ИС.

## 3 Проектирование от бизнес-архитектуры

В МАК именно бизнес-архитектура является отправной точкой для проектирования ИС. Под бизнес-архитектурой в контексте проектирования ИС понимается множество моделей автоматизируемых бизнес-процессов предприятия. В контексте данной работы рассматривается целевая бизнес-архитектура, т.е. модели бизнес-процессов с учетом использования ИС. Задача состоит в определении полного набора автоматизируемых функций бизнес-процесса.

### 3.1 Понятия бизнес-архитектуры

Архитектурные модели бизнес-архитектуры строится на трех основных элементах (Рисунок 2), определяемых в процедуре декомпозиции:

– бизнес-процесс, понимаемый в обычном смысле, как действие, в котором на основе одного или более видов входных объектов [ресурсов] создается ценный для клиента результат; бизнес-процесс может быть представлен комплексом действий, определяемых на основе его декомпозиции;
– бизнес-функция – это действие из комплекса действий, определяемых декомпозицией бизнес-процесса;
– бизнес-операция – это не декомпозируемая в рамках описания бизнес-процесса бизнес-функция, исполнителем которой является конкретный сотрудник.

Декомпозиция определяет явные связи между этими архитектурными элементами: бизнес-процесс декомпозируется на бизнес-функции, бизнес-функции, в свою очередь, декомпозируются на бизнес-функции и бизнес-операции. Процесс завершается, когда все бизнес-функции декомпозированы на бизнес-операции.

Бизнес-операция — это место, в котором возникает взаимодействие пользователя с ИС, и соответственно, в явном или неявном виде включает в себя автоматизируемые функции, поддерживающие ее.

### 3.2 Сервисы бизнес-операций

Бизнес-архитектура определяет функциональные требования пользователя к ИС. Описание бизнес-операций позволяет определить функции, которые должны быть автоматизированы. Набор этих функций, определенный для бизнес-операции, составляет содержание сервиса бизнес-операции (далее по тексту – сервис операции). Сервисы операций формируются для каждой бизнес-операции, включающей автоматизируемые функции. Они представляют собой архитектурные элементы, имеющие абстрактный характер, не предполагающий какой-либо самостоятельной реализации.

Сервис операции является связующим элементом, определяющим бесшовную связь между бизнес-архитектурой и функциональной архитектурой. Рисунок 2 демонстрирует эту связь: сервис операции декомпозируется на диалоги ИС, которые являются основным элементом функциональной архитектуры, обеспечивающим доступ к функциям ИС. Автоматизируемые функции бизнес-операции реализуются в диалогах.

---

[4] Связь между элементами архитектурных моделей, находящихся на разных уровнях абстракции.

Кроме этого, сервис операции представляет собой границу между бизнес-архитектурой и собственно архитектурой ИС. Он формализует требования к автоматизации бизнес-операций и служит основой для формирования технического задания на создание или развитие ИС.

**4 Функциональная архитектура**

Под функциональной архитектурой понимается архитектурное представление, включающее архитектурные модели структуры и состава функциональных компонент ИС, обеспечивающих доступ к «внутренним» функциям ИС, реализующим автоматизируемые функции. Иными словами, функциональная архитектура моделирует взаимодействие ИС с пользователями, а также с иными внешними агентами. Функциональная архитектура формирует облик ИС на основе представления композиций диалогов системы, определяет требования к диалогам и специфицирует интерфейсы с внешними ИС.

Диалог – это любой акт взаимодействия агентов, вызывающий изменение состояния ИС посредством запуска соответствующих программных компонент [19]. Таким образом, диалог понимается в широком смысле – это не только взаимодействие пользователя с компьютером, но также обмен сообщениями между любыми объектами ИС и внешними агентами.

Структура описания диалога (Рисунок 2) включает следующие архитектурные элементы:

– Функция уровня представления ИС (функция представления) – это функция ИС, реализуемая в диалоге, обеспечивающая возможность доступа к «внутренним» функциям ИС. Иными словами, через функции представления пользователь взаимодействует с ИС.
– Исходный ресурс и Целевой продукт – это информационные объекты соответственно, поступающие на вход диалога при его инициировании и формируемые в результате диалога на выходе из него.
– Форма диалога – представление диалога для пользователя, если в качестве агента предполагается пользователь.

Определяемые в диалоге функции представления, имеют следующую классификацию:

– реализация предусловий диалога при инициировании;
– ввод/вывод значений полей данных;
– обработка управляющих элементов диалога, изменяющих состояние ИС;
– реакция на ошибки и нештатные ситуации в ходе диалога;
– реализация постусловий при завершении диалога.

Функциональная архитектура детализируется в компонентной архитектуре. Связующим элементом, определяющим бесшовную связь, являются функции представления, которые декомпозируются на программные модули (Рисунок 3). Детальное рассмотрение компонентной архитектуры приводится в следующем разделе.

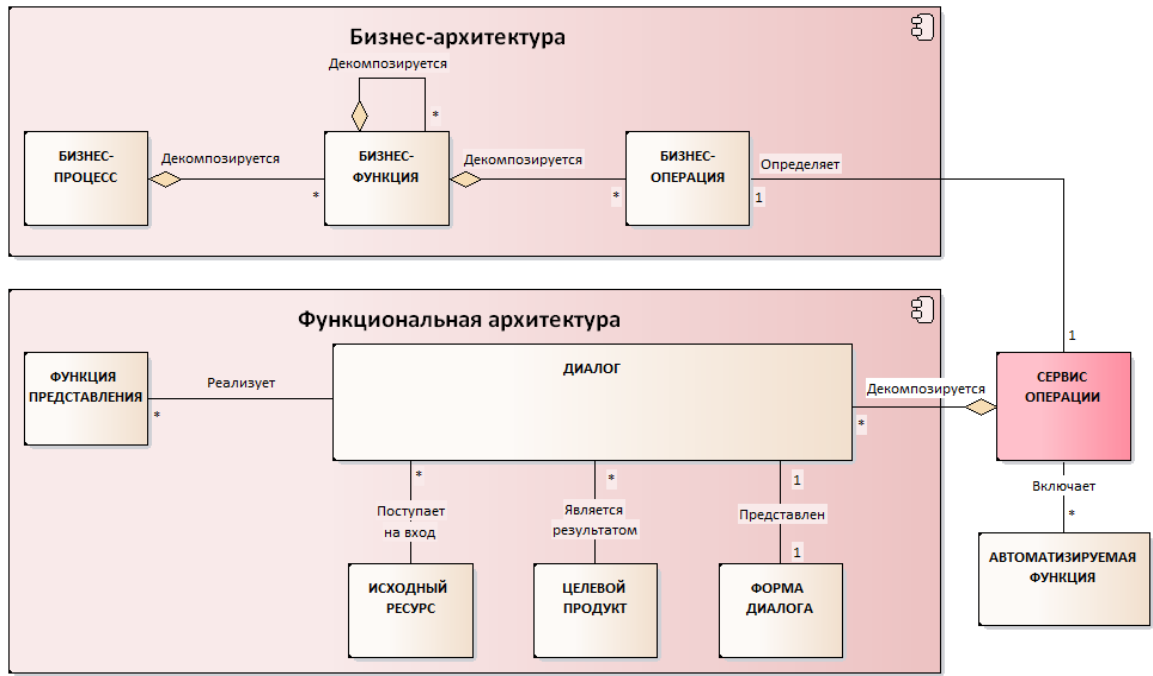

Рисунок 2 - Связь бизнес-архитектуры и функциональной архитектуры

## 5 Компонентная архитектура

Полученные в результате проектирования диалогов функции представления не учитывают компонентную архитектуру системы. Эти функции касаются лишь уровня представления ИС и не затрагивают ее внутреннее устройство. Представление архитектуры ИС в виде взаимодействующих функциональных компонент (подсистем и внешних ИС) позволяет выяснить, каким образом исполняются функции представления, определить внутренние функции логики приложения и управления данными.

Компонентная архитектура ИС устанавливает состав и взаимодействие функциональных компонент ИС, определяет программные модули и их распределение по функциональным компонентам ИС, уточняет функциональные требования к ИС.

Основными элементами модели компонентной архитектуры являются (Рисунок 3):

– функциональные компоненты (подсистемы и внешние ИС), определяемые выбором архитектурного шаблона (или композиция архитектурных шаблонов) и внешним окружением системы;
– программные модули, представляющие собой реализуемые в программном коде структурные части функциональных компонент ИС, вызываемые функциями представления.

Проектирование компонентной архитектуры начинается с определения компонентной структуры ИС, на основании которого система представляется множеством функциональных компонент (подсистем и внешних ИС). Модульный состав функциональных компонент системы определяется декомпозицией связующих элементов функциональной архитектуры – функций представления ИС, выполняемой на выбранной компонентной структуре. В декомпозицию функции представления могут входить ранее определенные модули для повторного использования, т.е. между функциями представления и модулями системы существует связь «многие ко многим».

В МАК для определения модульного состава ИС применяются диаграммы последовательности (*Sequence Diagrams*), формируемые для функций представления каждого диалога, описанного на уровне функциональной архитектуры. В качестве линий жизни на диаграммах используются функциональные компоненты. Таким образом, проектирование позволяет определить полный набор программных модулей для всех функциональных компонент.

## 6 Определение методов классов

МАК предлагает технологию разработки архитектуры данных (см. [23] и [24]), но изложение этой технологии выходит за рамки данной работы. Мы исходим из того, что архитектура данных тем или иным способом разработана на уровне ER-модели и включает все необходимые атрибуты, а сущности распределены по элементам компонентной модели. Задачей является определение методов классов архитектуры данных на основе сформированной модульной структуры: для каждого модуля компонентной модели определяется порядок его реализации в рамках объектно-ориентированной модели ИС.

Таким образом, связующим элементом, определяющим бесшовный переход от компонентной архитектуры к архитектуре данных, являются программные модули компонентной модели. Методы классов определяются декомпозицией каждого модуля на представленной архитектуре данных (Рисунок 3). Также, как и в случае компонентной модели, между модулями и методами классов существует связь «многие ко многим», соответствующая повторному использованию методов для реализации модулей. Определение методов классов также выполняется с помощью диаграмм последовательности, в которых линиям жизни соответствуют классы модели данных.

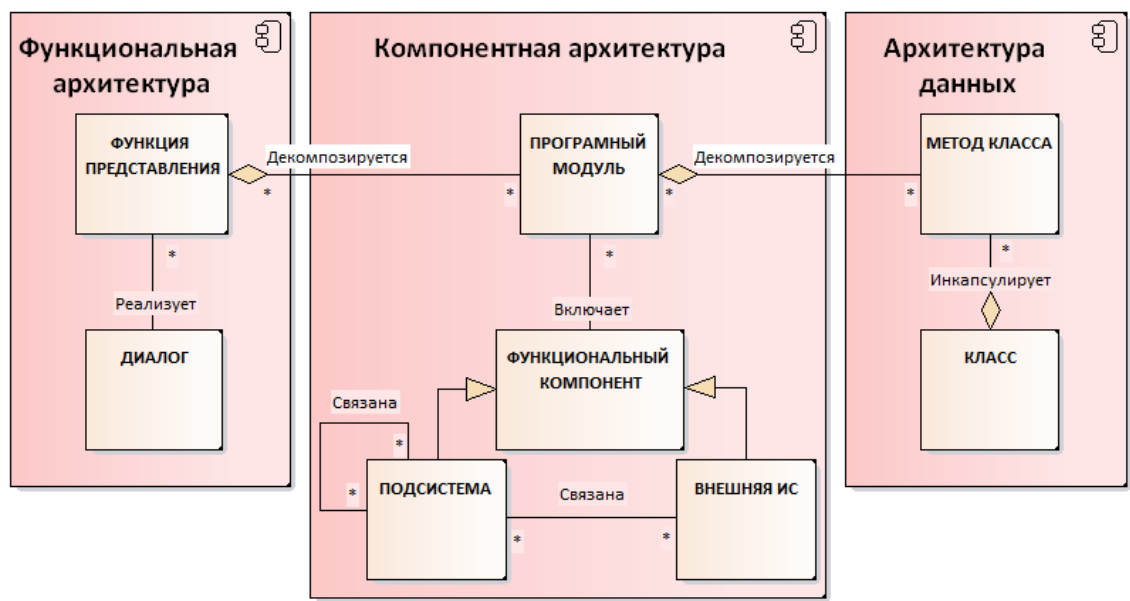

Рисунок 3 – Функциональная, компонентная архитектуры, архитектура данных и их связь между собой

## 7 Технологическая архитектура

Технологическая архитектура – это модель, представляющая техническую инфраструктуру ИС, включая решения в области вычислительной и телекоммуникационной инфраструктуры.

Технологическая архитектура, в контексте данной работы, представляет распределение элементов компонентной архитектуры (развертывание) по различным аппаратно-программным средствам и определяет необходимые интерфейсы между объектами технической архитектуры. Выбор архитектурных решений для технологической архитектуры ограничивают системные требования, атрибуты качества и требования к внешним интерфейсам.

С точки зрения уровня абстракции, технологическая архитектура выше компонентной. Рисунок 4 показывает, что связующим элементом между этими архитектурами служит элемент технологической архитектуры, соответствующий аппаратно-программному средству. В качестве декомпозиционной связи здесь выступает связь, отражающая расположение функциональных компонент на программно-аппаратных средствах ИС.

Архитектурное представление технологической архитектуры, как правило, выполняется с помощью диаграмм развертывания (*Deployment Diagrams*). Описание объектов диаграмм развертывания может включать либо требования к характеристикам аппаратно-программного средства, либо указание на конкретную модель устройства.

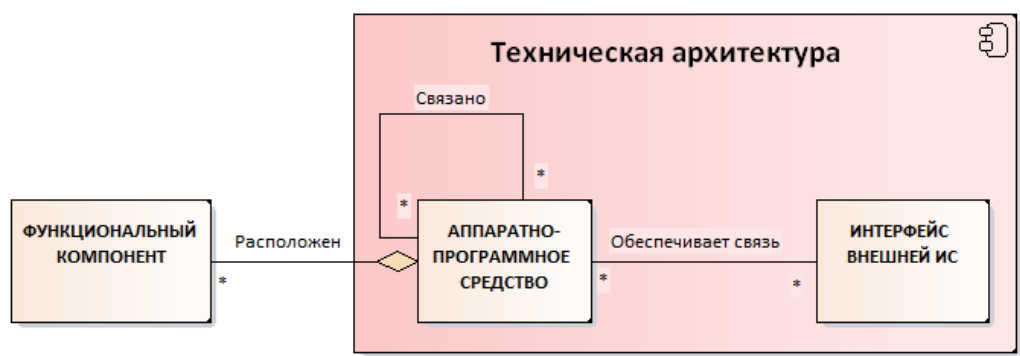

Рисунок 4 - Связь компонентной и технической архитектур

## 8 Заключение

Рассмотренный в данной работе метод обеспечивает бесшовный переход от бизнес-архитектуры через декомпозицию сервиса операций к функциональной архитектуре ИС, определяющей диалоги системы и функции представления. Функции представления декомпозируются на программные модули компонентной архитектуры и, наконец, модули декомпозируются на методы классов архитектуры данных. Иными словами, каждое архитектурное представление выводится из архитектуры предыдущего уровня абстракции, определяя единую архитектуру ИС.

При таком подходе обеспечивается полнота функциональной реализации ИС, поскольку декомпозиция бизнес-процесса позволяет точно сформулировать требования пользователя к автоматизируемым функциям, а также сформулировать концепцию создаваемой ИС и направления ее развития на основе определения полного множества бизнес-операций.

Дальнейшее проектирование функциональных компонент на различных уровнях абстрактного описания ИС по сути является обоснованным выводом необходимой и достаточной функциональности ИС, что позволяет избежать ошибок, связанных с технологическими разрывами между архитектурными моделями, недостаточной или избыточной функциональностью разрабатываемой ИС.

К преимуществам, которые показал МАК в реальном проектировании и сопровождении ИС, можно отнести обеспечение прозрачности соответствия результата проектирования предъявляемым заказчиком требованиям за счет упрощения процессов валидации и верификации. Наличие трассировки между элементами архитектурных моделей позволяет осуществлять быструю локализацию необходимых изменений и доработок для выпуска новых релизов и версий ИС. Регламентация процесса разработки, связь архитектурного описания с артефактами с учетом возможности генерации проектных и эксплуатационных документов на основе архитектурных моделей [21] обеспечивает значительное сокращение сроков проектирования и внесения необходимых изменений в ИС.

В заключение хотелось бы обозначить темы, которые не нашли отражения в этой работе. Дальнейшего исследования требуют вопросы соотношения МАК и сервис-ориентированной архитектуры (СОА), а именно вопросы идентификации сервисов и реализации компонентной архитектуры ИС на шаблоне СОА. Вообще, работа с внешними информационными ресурсами, будь то веб-сервисы или библиотечные классы, требует осознания в контексте МАК.